\documentclass[aps,prl,showpacs,floats,twocolumn,amsfonts,amssymb,unsortedaddress,floatfix]{revtex4}
\input epsf
\usepackage{bm}
\usepackage{amsmath}
\usepackage{graphicx}

\begin{document}

\newcommand{\ud}{\mathrm{d}}
\renewcommand{\bottomfraction}{0.7}
\renewcommand{\topfraction}{0.7}
\renewcommand{\textfraction}{0.2}
\renewcommand{\floatpagefraction}{0.7}
\renewcommand{\thesection}{\arabic{section}}

\addtolength{\topmargin}{10pt}


\newcommand{\eq}{\begin{equation}}
\newcommand{\en}{\end{equation}}

\newcommand{\eqa}{\begin{eqnarray}}
\newcommand{\ena}{\end{eqnarray}}

\newcommand{\eqan}{\begin{eqnarray*}}
\newcommand{\enan}{\end{eqnarray*}}

\newcommand{\spz}{\hspace{0.7cm}}
\newcommand{\lbl}{\label}


\def\Bbb{\mathbb}


\newcommand{\Dslash}{{\slash{\kern -0.5em}\partial}}
\newcommand{\Aslash}{{\slash{\kern -0.5em}A}}

\def\d{\partial}
\def\({\left(}
\def\){\right)}
\def\zbar{\overline{z}}
\def\sqr#1#2{{\vcenter{\hrule height.#2pt
     \hbox{\vrule width.#2pt height#1pt \kern#1pt
        \vrule width.#2pt}
     \hrule height.#2pt}}}
\def\smallsquare{\mathchoice\sqr34\sqr34\sqr{2.1}3\sqr{1.5}3}
\def\square{\mathchoice\sqr68\sqr68\sqr{4.2}6\sqr{3.0}6}
 
\def\thinspace{\kern .16667em}
\def\punto{\thinspace .\thinspace}
 
\def\xp{x_{{\kern -.2em}_\perp}}
\def\subp{_{{\kern -.2em}_\perp}}
\def\kperp{k\subp}

\def\derpp#1#2{{\partial #1\over\partial #2}}
\def\derp#1{{\partial~\over\partial #1}}

\def\zbar{\overline{z}}
\def\wbar{\overline{w}}
\def\dbar{\overline{\partial}}
\def\ez{{{\bf e}}_z}
\def\ezbar{{{\bf e}}_{\zbar}}
\def\vF{ v_{_{\rm F}} }
\def\EF{ E_{_{\rm F}} }
\def\lB{ \ell_{\rm B} }
\def\orbell{ \ell }
\def\Kp{\bm{K}_+}
\def\Km{\bm{K}_-}
\def\psiA{\psi_{_{A}}}
\def\psiB{\psi_{_{B}}}
\def\defeq{:{\kern -0.5em}=}

\title{Thermal Stability of Strained Nanowires}

\author{Cristiano~Nisoli$^{1}$, Douglas Abraham$^{1,2}$, Turab Lookman$^{1}$ and Avadh Saxena$^{1}$}
\affiliation{$^{1}$ T-Division and Center for Nonlinear Studies, Los Alamos National Lab, Los Alamos NM 87545 USA \\
$^{2}$ Rudolf Peierls Centre for Theoretical Physics, 1 Keble Road Oxford, OX1 3NP England }

\date{\today}
\begin{abstract}
Stranski-Krastanov strained islands undergo a shape anisotropy transition as they grow in size, finally evolving toward nanowires. This  effect has been explained until now via simple energetic models that neglect thermodynamics. We investigate theoretically the stability of strained nanowires under thermal fluctuations of the long side.  We find phase transitions from nanowires back to nanoislands as the temperature is increased and as the height of the nanostructure is raised or lowered and we predict regions of phase coexistence. Our results are general, but explain recent data on the growth of erbium silicide on a vicinal Si surface. 
   
\end{abstract}


\maketitle

The investigation of two-dimensional nanoislands has provided fundamental understanding of the mechanism of epitaxial growth.  Instabilities in heterostructure  growth are exploited to form self-organized nanodots and nanowires that  are of much  smaller size than similar nanostructures realized on lithographically pre-patterned substrates~\cite{Deng, Xu}. Typically,  competition of plastic and elastic relaxation processes in  strained epitaxial layers drives their formation~\cite{Stangl, Vanderbilt}. In Stranski-Krastanov growth~\cite{Stangl, Stranski}, where interactions between adatoms and of adatoms with the surface are comparable, both dislocated or single crystalline islands can grow after the initial wetting layer has reached a critical thickness. Formation of single crystalline rather than dislocated islands depends on the material, the amplitude of the lattice misfit and the possible interplay of the Asaro-Tiller-Grinfield instability~\cite{Asaro,Grinfeld}. It can be controlled in many ways, e.g. by undulations and/or patterning in the surface layers of the substrate~\cite{Chiu} or by tuning areal island densities and temperature~\cite{Liang}. 

It has been predicted and observed experimentally that dislocation free islands in  early stages of growth can undergo  a shape anisotropy transition as their size becomes critical~\cite{Tersoff, Li, Zandvliet}. Eventually their  growth becomes quasi-one dimensional and they form  perfect nano-objects with well defined height and width, that is, nanowires. Previous theoretical treatments have been based on simple energetic models  of the island/wire that include competition between strain relief and surface energy.  Although useful, these models completely neglect thermal fluctuations which are important in such quasi-one dimensional systems because high entropic gains of the (thermodynamically extensive) long sides can  lead to various instabilities. 

Our objective is thus to introduce  thermal effects in the theory within the context of a statistical mechanical model.  We find that above a critical temperature the wire does indeed become unstable and this temperature depends on the threshold of the plastic relaxation of the film. Moreover, we find that the wires  become unstable as the deposition continues at constant temperature. 
We explain the observations of a recent set of experiments on the growth  of erbium silicide  nanostructures on the vicinal Si(001) surface that was studied at different post-annealing temperatures and coverages~\cite{Zhu,Zhou,Ji}.

\begin{figure}[b!]
\includegraphics[width=3.0 in]{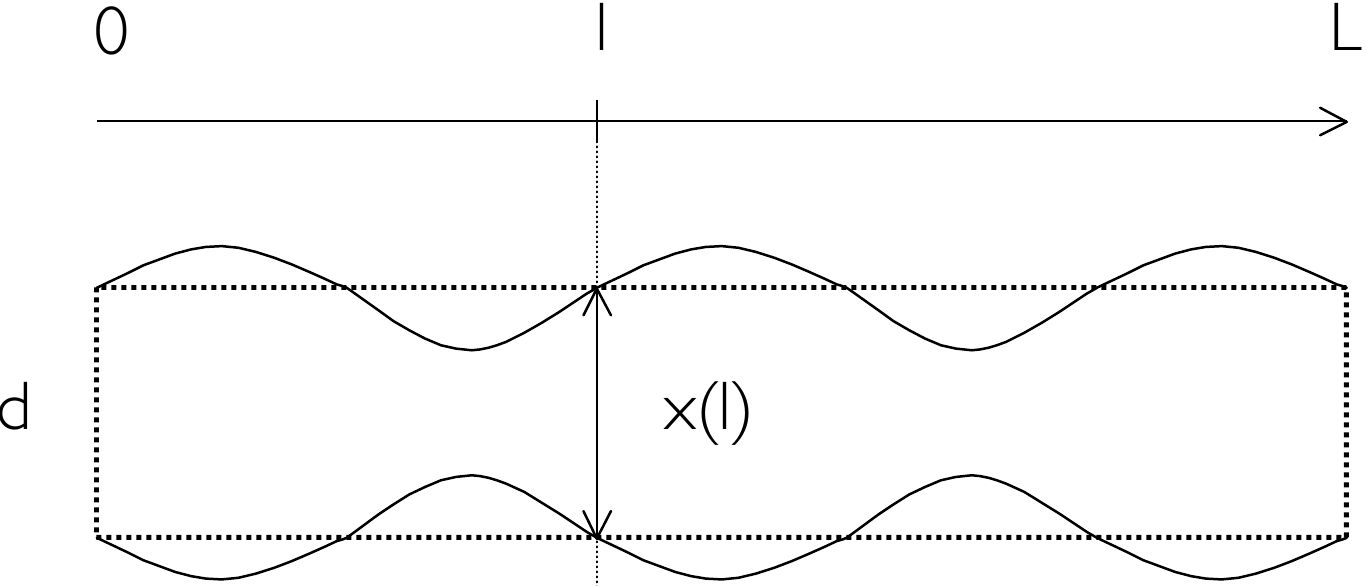}
\caption{The nano-wire with lateral edges $x_1(l)$, $x_2(l)$ allowed to fluctuate in the coordinate $l$, $0\le l \le L$; $x(l)=x_1(l)-x_2(l)$ is the relevant degree of freedom for our problem. Amplitude of fluctuations is largely exaggerated in figure.}
\label{Fig0}
\end{figure}

We first review  the Tersoff and Tromp  model of the shape transition~\cite{Tersoff} on which more detailed studies have been based  ~\cite{Li, Zandvliet}. By considering both surface energy and strain relaxation 
Ref~\cite{Tersoff} arrives at the following expression for the energy for unit area of a rectangular island of dimensions $L$, $X$ 
\begin{equation}
\frac{\epsilon}{\epsilon_o}= -\frac{d}{X} \ln \frac{ X e}{d}  -\frac{d}{L} \ln \frac{ L e}{d};
\label{en}
\end{equation}
the parameters $d$, $\epsilon_o$ depend  upon surface energy and strain. The superficial density of energy  in (\ref{en}) is minimized by a square island of size $X=L=d$ and energy $d^2 \epsilon_o$.  
Yet, depending on the growth condition, islands can grow metastably larger than the optimal size $d^2$.  Minimizing at fixed area  $A=XL$, Tersoff and Tromp obtained a critical area, $A_c=(e d)^2$,  below which  symmetric shapes, or $X=L=\sqrt{A}$, are energetically favorable and above which  a symmetry breaking occurs. That is, as the surface $A$ increases beyond the critical value, one of the lateral sizes,  say  X, shrinks down from $e d$ asymptotically  approaching  the  equilibrium value $d$. The other side keeps growing eventually linearly as $L\sim A/d$. Indeed this asymptotic behavior can be  extracted from (\ref{en}) by taking the limit  $L\rightarrow \infty$ and then obtaining the superficial energy for lateral growth
\begin{equation}
V(x)= - \epsilon_o\frac{d}{x} \ln\frac{x e}{d},
\label{V}
\end{equation} 
which has as expected  minimum in $x=d$ (we will from now on use lower case $x$ to denote the finite edge, and upper case $L$ for the long edge - ``long'' as in the thermodynamic limit, Fig.~\ref{Fig0}). 

We will explore the thermal stability of a long wire whose lateral size $d$ minimizes the asymptotic  energy in (\ref{V}) under fluctuations of its long edges.  We allow  the lateral edges $x_1(l)$, $x_2(l)$ to fluctuate in the coordinate $0\leq l \leq L$. The problem is simplified by the introduction of $x=x_1-x_2$, $x_{+}=x_1+x_2$. The elastic energy of those fluctuations is thus simply  $T= k x'^2_1+ k x'^2_2= \frac{1}{2}  k x'^2+\frac{1}{2} k x'^2_{+}$. As the term in $x_+$ only contributes to equipartition, it can be disregarded.
 The partition function  in the continuum limit is a gaussian path integral over the allowed fluctuations
\begin{equation}
Z\left(\beta\right)=\int \prod_l \left[\mathrm{d}x\left(l\right)\right] e^{-\beta \int_0^L \mathrm{d}l\left[\frac{ k }{2}x'^2(l)+  d V\left(x\left(l\right)\right)\right]} \!\ .
\end{equation}
Standard transformations~\cite{Kleinert, Feynman} allow us to write it as the trace
\begin{equation} 
Z\left(\beta\right)=\left(\frac{2\pi}{\beta k \lambda}\right)^{L \lambda} \mathrm{Tr}\left[ \exp{\left(-\frac{L}{d}  \hat{H}_{\beta}\right)}\right]
\label{trace}
\end{equation}
of the one-particle, temperature dependent hamiltonian given by
\begin{equation}
\hat{H}_{\beta}=-\frac{d}{2 \beta k }\frac{\mathrm{d^2}}{\mathrm{d}x^2 }+\beta d^2 V(x).
\label{schrodinger}
\end{equation}
The coefficient in front of the trace simply adds to the free energy the equipartitive contribution of free oscillators  ($\lambda$ is the linear density of oscillators   $\lambda= K/k$, with $K$ the microscopic spring constant). We thus only need to concern ourselves with the diagonalization of the (temperature dependent) effective Hamiltonian.
In the limit of a long wire, or $L/d \to \infty$, the density operator in (\ref{trace}) projects on the lowest bound state -- provided that $\hat{H}_{\beta}$ has one. One can reproduce the previous calculation for the partition function and demonstrate that the probability of a lateral size $x$ is given by
$p(x)=\lim_{J \to 0^+} |\psi_{\beta}^{0,J}(x)|^2$ where $\psi_{\beta}^{0,J}$ is the lowest bound state of the operator $\hat{H}_{\beta}+J x$, where $J$ is a chemical potential.  Clearly when  $\hat{H}_{\beta}$ has no bound states -- which might happen for $\beta$ below a certain $\beta_c$ -- then $p(x)$ is flat, and we conclude that the elongated structure is not stable at that temperature. Thus the search for stability of long wires is reduced to the problem of  existence of a bound state in a suitable 1-particle hamiltonian operator,  a situation reminiscent of localization/delocalization of one-dimensional interfaces of the solid on solid model \cite{Abraham, Burkhardt, Burkhardt2}. 

Before exploring this transition, one can immediately show a relationship between the critical exponent of $\bar x=\int x  |\psi_{\beta}^{0}(x)|^2\mathrm{d}x\sim t^{-\beta}$ and that of the specific heat $C\sim t^{-\alpha}$ when  $t=(T-T_c)/T_c \to 0^-$. Indeed, from (\ref{trace}) we have $F\simeq TL E^0/d$,  while for $T\sim T_c^-$ one has $ E^0\sim d(2\beta k)^{-1}\bar x ^{-2}$, since it is the exponential tail of the bound state around infinity that dictates $\bar x$. Thus $f\sim t^{2\beta}$, or
$\alpha=2(1-\beta)$. We show at the end that $\beta=1$.

\begin{figure}[t!]
\includegraphics[width=3 in]{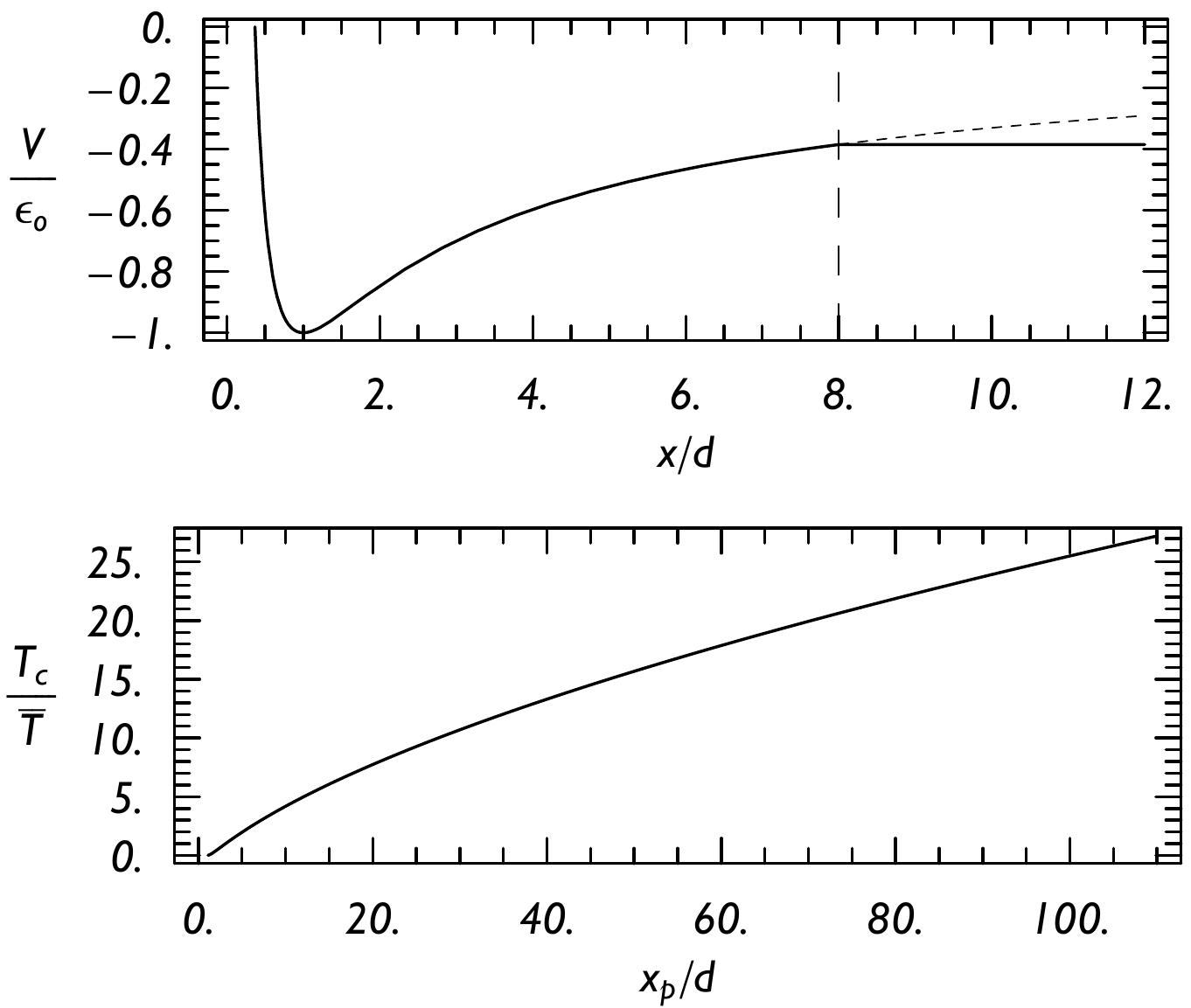}
\caption{From top to bottom:  the potential of (\ref{V}) (solid line) with plastic limit set at $x_p=8 d$ (dashed line). Behavior of  $T_c$ (in units of $\bar T=\pi^{-1}\sqrt{2^3 k \epsilon_od^3 }$) as a function of the plastic limit $x_p$}
\label{Fig1}
\end{figure}

The potential in (\ref{V}) converges to zero slower than $1/x$ at infinity while at zero it  goes to infinity slower than the  centrifugal energy; thus $H_{\beta}$, for any value of coupling constant (i.e. any temperature) has at least as many bound states (excluding S waves) as the hydrogen atom: infinite. Such a cursory look might convince the reader that nanowire formation should always be stable with respect to thermal fluctuations. In fact it is the long range behavior of the potential  that commands possible instability, but  (\ref{V}) is clearly unphysical for large $x$; simple intuition suggests that for large islands, energy density must be constant, as plastic relaxation of the elastic strain becomes unavoidable when the islands grow in size. Relaxation via dislocations has been observed experimentally in nanoislands and has already found to promote many complex phenomena~\cite{LeGoues}, only recently explored via simplified approaches~\cite{Johnson,Spencer}. 
Fortunately our scenario is  simpler: islands grow in a  single crystalline manner until they reach a threshold size in which dislocations appear. Once dislocations start to form, they can coalesce, move and make other dislocations more likely. It is thus both reasonable and simple to introduce  an ad hoc plastic limit $x_p$ beyond which the  energy of the island simply grows linear with size.

\begin{figure}[t!]
\begin{center}
\vspace{1mm}\includegraphics[width=3.3 in]{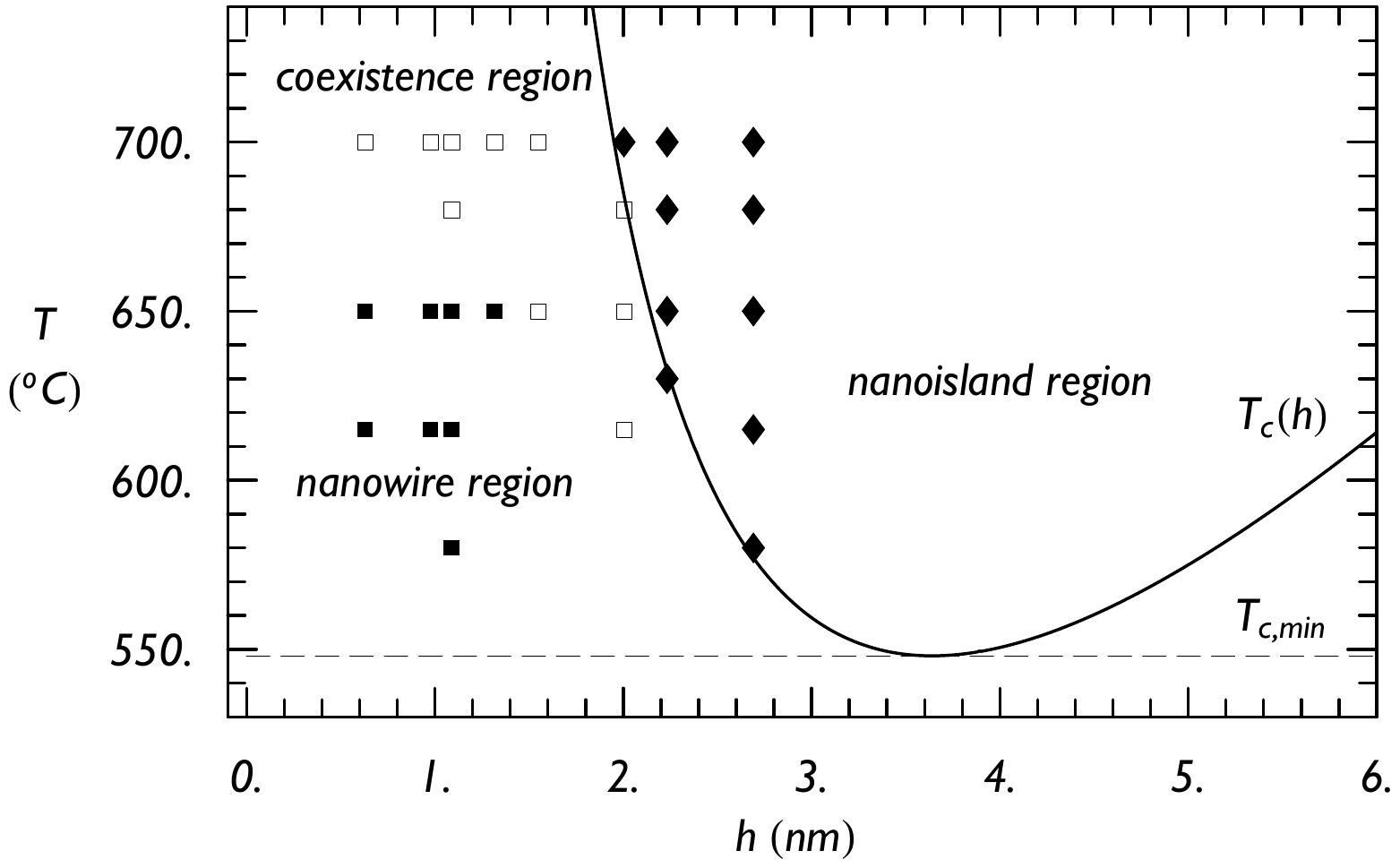}\vspace{-3 mm}
\end{center}
\caption{Phase diagram $(T,h)$ from (\ref{phase}) and the experimental findings of Zhou et al.~\cite{Zhou}. Boxes denote nanowires,  empty boxes regions of coexistence while diamonds correspond to nanoislands (the relative coverage is 0.29,   0.73,     0.87,     1.16,     1.45,     2.03,     2.32,     2.90 Mono Layer or ML). We have chosen our parameters  arbitrarily (yet reasonably) to provide a decent superposition with Cai's results:  $\frac{\Gamma_p}{2 c }=6$ nm, $\frac{\Gamma}{2c}=1$ nm,  $T_{c,\mathrm{min}}=548~ ^o$C. Note that for low $h$, where $T_c$ diverges, the region of coexistence of islands and wires is still experimentally accessible, as discussed theoretically in the text. Note also the region of nanowire stability with no transition for $T<T_{c,\mathrm{min}}$ and for large heights $h>3.5$ nm.}
\label{Figphase}
\end{figure}


\begin{figure}[t]
\includegraphics[width=3 in]{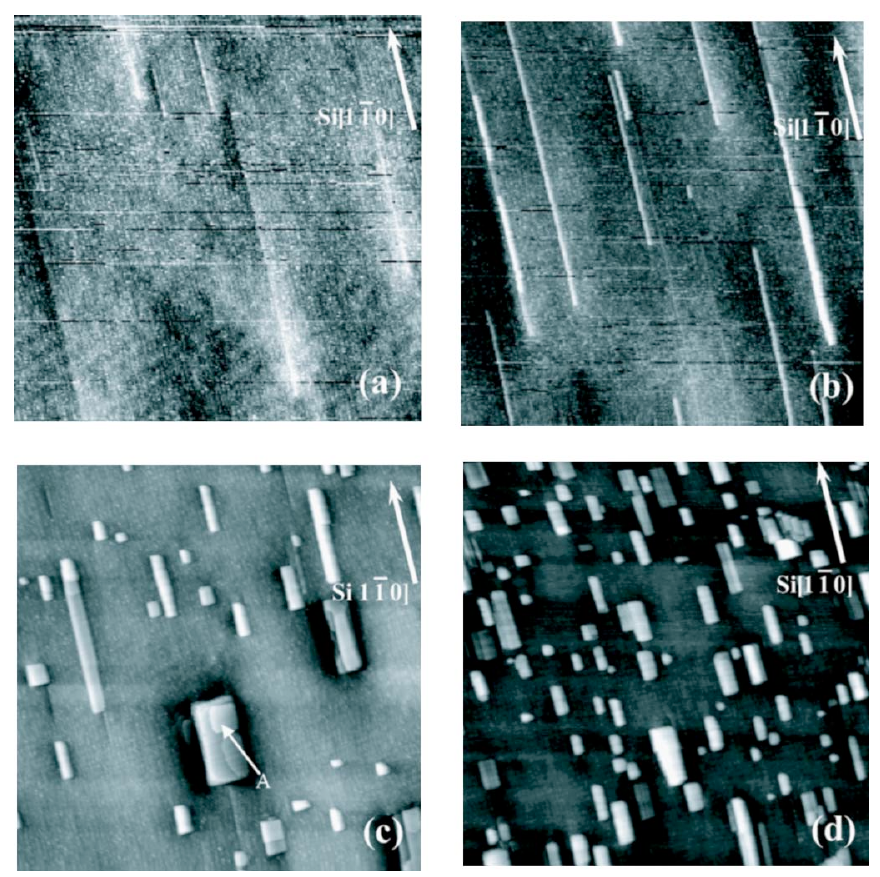}\vspace{-2 mm}
\caption{ A series of STM images (all 300 X 300 nm$^2$) taken on the vicinal Si(001) surface with different Er coverages. The samples are post-annealed  at 650¡ C for 10 min.  (a) 0.29 ML. (b) 0.73 ML.(c) 2.03 ML. (d) 2.90 ML. From ref~\cite{Zhu}.}
\label{FigC}
\end{figure}

 We substitute the potential in (\ref{V}), $V \to V_p$, with $V_p(x)=V(x)$ when $x < x_p$ and $V_p(x)=V(x_p)=V_p$ for $x>x_p$.  $V_p(x)$ is now a well for $x>0$ (Fig.~\ref{Fig1}) and thus might or might not admit a bound state, depending on the value of $\beta$.  The Schr\"odinger problem  for the potential in (\ref{V}) has not been solved yet, nevertheless the critical temperature corresponding to the disappearence of the lower bound eigenstate can be obtained via a WKB approximation, which works well  when $T_c$ is not too large~\cite{Fermi}:
\begin{equation}
T_c=\frac{2^{3/2}}{\pi}\sqrt{k \epsilon_od^3 }g\left(\frac{x_p}{d}\right).
\label{Tc}
\end{equation}
For $\xi>1$, $g(\xi)$ is defined  by
\begin{equation}
g(\xi)=\int^{\xi}_{\xi_0}\sqrt{\frac{1}{s} \ln{(e s )}-\frac{1}{\xi} \ln{(e\xi )}}~\mathrm{d}s
\label{geq}
\end{equation}
 $\xi_0$ is the lower zero of the integrand. 

Equations (\ref{Tc}), (\ref{geq}) link critical temperature to the plastic limit. Figure~\ref{Fig1} depicts  the behavior of $T$ as a function of the threshold for dislocation $x_p$ assuming all the other parameters are constant.  One would deduce that  in general stability can be achieved at higher temperatures by pushing higher the single crystalline threshold. Yet the phase transition was  obtained in the limit  $d/L\downarrow 0$. Consider now (\ref{trace}), (\ref{schrodinger}) for $L/d$ large but finite. For $T>T_c$, where $T_c$ is the critical temperature, there are clearly no nano-wires, as there are still no bound states. When $T<T_c$, the operator in (\ref{schrodinger}) has (at least) one bound state, and yet the density operator does not just project on the ground state: it leaves some probability $p\propto e^{-\frac{L}{d}|E_{\beta}^o|}$ of a jump from the bound state to the continuum spectrum, a jump activated by the small ``pseudo-temperature'' $d/L$. In the proximity of  $T_c$,  we obtain (see later) $E_{\beta}^o=\pi^2 \epsilon_o d^2 (T_c-T)^2/2 T_c^3$; $p$ is thus a gaussian in $T-T_c$ of relative width 
\begin{equation}
\frac{\Delta T_c}{T_c}= \frac{1}{\pi}\sqrt{\frac{T_c}{\epsilon_o L d}}~,
\label{spread}
\end{equation}
which implies a gray zone of coexistence of both islands and wires of the order of $10^{-1} T_c$ (if we assume $k $ of the same order as $ \epsilon_o d$, $L/d\sim 10^2$ and $x_p/d\sim10-100$). Notice that the width in (\ref{spread}) increases with $T_c$. That makes it possible for a region of coexistence of islands and wires to be accessible even when $T_c$ is beyond practical reach. This seems to be the case in ref~\cite{Zhou}, as explained later. 

It is interesting to explore the behavior of the critical temperature with deposition.  In general the plastic limit as well as  $\epsilon_o$, $k$ and $d$ depend on island height $h$, surface energy and strain.  Tersoff and Tromp found $d/e= \phi h \exp(\Gamma/2ch)$ and $\epsilon_o =2c h/d$, where $\Gamma$ is proportional to the surface energy, $c$ to the strain energy and $\phi$ is a number which depends on the contact angle~\cite{Tersoff}. Clearly, the linear elastic constant $k=\chi h$ is also proportional to the island height. In the simpler approach, the energy balance $x_p(V(x_p)-V(x_p/2))=\Gamma_p$ provides a  criterion for the formation of a defect of energy (per unit length) $\Gamma_p$, which gives
$ x_p/d
=4 \exp\left(\Gamma_p/2 c h-1\right)
$.
So, if  we neglect the dependance of  $\Gamma_p$  upon temperature we can relate critical temperature with the island height and thus obtain the critical line
\begin{equation}
T_c(h)=\frac{2 e \phi}{\pi}\sqrt{\chi c }~ h^2 e^{\frac{\Gamma}{2ch}}g\left(\frac{4}{e} e^{\frac{\Gamma_p}{2 c h}}\right).
\label{phase}
\end{equation}
In Fig.~\ref{Figphase} we plot (\ref{phase})  as a phase diagram for a nanowire/nanoisland under the above assumptions, along with experimental results of Zhou {\it et al.}~\cite{Zhou}. One sees  that below a certain temperature  $T_{c,\mathrm{min}}$ the growth of a nanowire by further deposition is always stable. Above $T_{c,\mathrm{min}}$, deposition will cause growth in height until  a critical value is encountered, above which the wire becomes unstable. 
Equation (\ref{phase}) also predicts a region of wire stability for higher values of $h$. As Fig.~\ref{Figphase} shows, the interval of instability in $h$ decreases in size with the temperature, until it disappears below $T_{c,\mathrm{min}}$. 
Depending on materials, kinetic parameters and other physical circumstances, only a portion of the phase diagram of Fig.~\ref{Figphase} might be accessible.

Our predictions explain  the experimental results of Cai's group on the self organization of erbium silicide on Si (001). They obtained  phase transitions from nanowires to nanoislands, both when increasing post-annealing temperatures at fixed coverage~\cite{Zhu} and when increasing  coverage (and thus the height of nanostructures) at fixed post-annealing temperatures as in Figs.~\ref{Figphase},~\ref{FigC}~\cite{Zhu,Zhou}. In a yet different approach, the same group was able to  control the efficiency of  the chemical synthesis of the silicide via different deposition techniques,  and could obtain nanostructures of different heights (2.0, 1.5 , 0.5 nm) at the same temperature: again transition from nanowires to nanoislands was observed as the height increased~\cite{Ji}, as well as the coexistence zone. 

Finally, let us tie some loose ends. The WKB method seems appropriate to compute the critical temperature but not for critical exponents. It is clearly so for a case that can be solved exactly, the square well. And yet, as the length scale diverges at the transition,  one expects the actual shape of the well to become irrelevant to the exponents. With the approximation  $g(\xi)\simeq (\xi-1)/2$, which works fine for $\xi$ of the order of 3-10, one finds that our critical temperature corresponds to that of a square well of depth $\epsilon_o$ and size $a$ such that $ 4 \epsilon_o a^2=\epsilon_o (x_p-d)^2$.  The problem for the square well potential can be solved easily and returns for the ground state $E_{\beta}^o=\frac{\pi^2}{2} \epsilon_o d^2 \frac{(T_c-T)^2}{T_c^3}$, and then $1/\bar x\sim t $ and thus $f\sim t^2 $, for $t<0$ while $f=0$ for $t>0$: $f$ has no kinks, hence entropy is continuous at the transition  and there is no latent heat, but a discontinuity in the specific heat.  One can check that on the contrary a WKB study of a square well (and also for our truncated well) returns different exponents, $1/\bar x\sim t^{1/2}$, and thus a latent heat.

In conclusion, we have studied  the  stability of nanowire fabrication under thermal fluctuations. We find phase transitions from wires to islands, both by increasing temperature and the  height of the nanostructure. Plasticity of extended films plays a crucial role in our analysis.  Our results show that the transition can be more or less sharp depending on the average length of the wires, and is preceded by regions of coexistence of wires and islands.  We thank Dr. Cai and her group for sharing figures and data. 

This work was carried out under the auspices of the National Nuclear Security Administration of the U.S. Department of Energy at Los Alamos National Laboratory under Contract No. DE-AC52-06NA25396.



\end{document}